\documentclass[12pt]{iopart}

\usepackage{iopams}  
\usepackage{graphicx}  
\begin{document}

\title[Phase transition 
between the quantum spin Hall and insulator phases in 3D]
{Phase transition 
between the quantum spin Hall and insulator phases in 3D:
emergence of a topological gapless phase}
\author{Shuichi Murakami}
\address{Department of Applied Physics, University of Tokyo,
Hongo, Bunkyo-ku, Tokyo 113-8656, Japan}

\ead{murakami@appi.t.u-tokyo.ac.jp}
\begin{abstract}
Phase transitions between the quantum spin Hall and the 
insulator phases in three dimensions are studied. 
We find that in inversion-asymmetric systems there 
appears a gapless phase between the quantum spin Hall and 
insulator phases in three dimensions,
which is in contrast with the two-dimensional case.
Existence of this 
gapless phase stems from a topological nature of gapless
points (diabolical points) in three dimensions, but not in two dimensions.
\end{abstract}

\submitto{\NJP}
\maketitle

\section{Introduction} 
In the intrinsic spin Hall effect (SHE) \cite{Murakami03a,Sinova04},
an external electric field applied to a doped semiconductor induces a
transverse spin current. 
It has been attracting special interest from the following aspects.
First, it 
can produce spin current without breaking the time-reversal symmetry 
(T-symmetry), namely 
without magnetism or
magnetic field, which 
may be potentially important for spintronics device applications.
Second, the mechanism itself is dissipationless and thus it may open 
a way for spintronics devices with less power consumption. 
Third, the SHE is driven by the spin-orbit couping, which can be
even larger than room temperature. Thus it 
is expected to survive even at room temperature, as has been confirmed 
experimentally \cite{Saitoh06,Stern06,Kimura07}.
In particular, recent experiments on platinum shows the largest spin
Hall conductivity so far \cite{Kimura07}, of about 240$\Omega^{-1}$
cm$^{-1}$ at room temperature. This might be attributed as an
intrinsic SHE caused by near-degeneracy near the 
Fermi energy \cite{Guo07a}.
 
Henceforth we restrict ourselves on time-reversal-symmetric systems.
In relation to the SHE in conducting systems, there have been a
growing interest in the SHE in insulating systems. The first proposal is 
spin Hall insulators \cite{Murakami04c}.
It was shown that in insulators such as
HgTe under uniaxial pressure or in PbTe the 
spin Hall conductivity is nonzero. In a sense, 
they are ordinary insulators with spin-orbit coupling.
Another proposal is the quantum spin Hall (QSH) systems, both 
in 2D \cite{Kane05a,Kane05b,Bernevig05a}
and in 3D \cite{Fu07a,Moore07}. 
They are 
insulators in the bulk while the boundaries (i.e. edges in 2D or 
surfaces in 3D) are gapless and carry spin currents. 
They can be regarded as topological insulators; 
these gapless boundary states are topologically protected against
T-symmetric perturbations \cite{Wu05,Xu05}. 
Experimental observations are yet to be made.
We recently proposed that thin-film bismuth is a good candidate for the 
2D QSH phase \cite{Murakami06b}. Another candidate for the
2D QSH phase is the CdTe/HgTe/CdTe
quantum well proposed \cite{Bernevig06f}. The QSH phases 
are, however, yet to be realized experimentally. 
Distinction between the QSH and ordinary insulating (i.e. spin Hall insulator)
phases 
is the absence or presence of (topologically protected) boundary states,
and is characterized by the $Z_2$ topological number of the bulk states
\cite{Kane05b,Fu06a,Fu06c}.

In the previous paper \cite{Murakami07a} we considered 
a phase transition between the QSH and the 
insulating phases in {\it two} dimensions.
It was found that the phase transitions are classified into
two cases, corresponding to 
the presence or absence of the inversion  symmetry (I-symmetry) 
in the systems.
It was also found that the effective theory describing the 
phase transition consists of two decoupled theories of two-component
fermions.

In this paper we study the phase transition between the two phases 
in {\it three} dimensions in the bulk. 
At first sight it might be similar to the two-dimensional case studied
in \cite{Murakami07a}.
We find in this paper that it is not. Topological properties 
such as the present problem can be very different for different
dimensions. 
In 3D, the inversion-symmetric case is similar
to that in the 2D. Nevertheless the inversion-asymmetric case is 
different; when we vary an external parameter which drives the 
phase transition, there appears a gapless phase between the QSH 
and the insulating phases. An existence of this gapless phase is enforced
from a topological origin, and this phase does not exist in 2D.
These discussions not only deepen our understanding toward the $Z_2$ 
topological numbers, but also imply some hints for materials search 
for the 3D QSH phases.

The paper is organized as follows. In Section 2 we discuss 
generic phase transitions between the two phases in 3D.
In Section 3 we discuss the implications of the present theory
for materials search for QSH phases. Section 4 is devoted for 
conclusions and discussions.
We neglect effects of interactions and impurities in the present
paper.

\section{Phase transitions between the quantum spin Hall and insulating
phases in 3D}
\subsection{Introduction on the quantum spin Hall phase and 
$Z_2$ topological number}
\begin{figure}[h]
\includegraphics[scale=0.6]{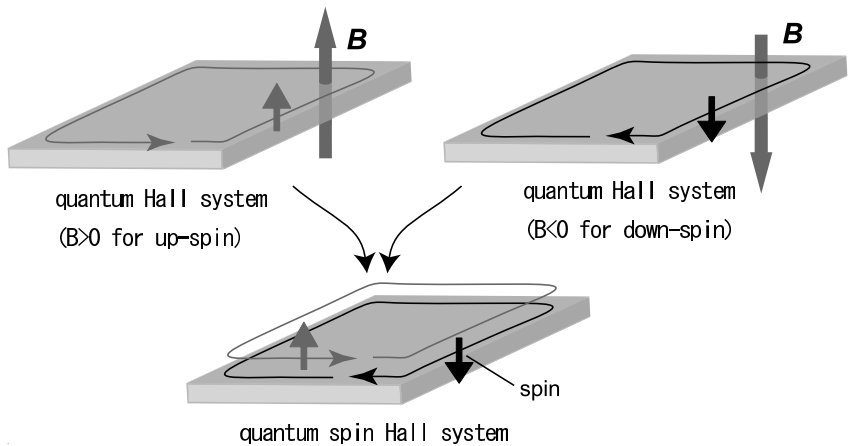}
\caption{Schematic picture of the quantum spin Hall system 
as a superposition of two quantum Hall systems.}
\label{fig:QSH}\end{figure}

We first review the QSH phase and 
$Z_2$ topological number $\nu$ \cite{Kane05a,Kane05b}, 
The simplest example of the QSH phase can be realized by 
a superposition of two quantum Hall systems for the up- and down-spins
having opposite (effective) magnetic field (see \fref{fig:QSH}). 
Suppose for the up-spin (down-spin) subsystem the quantum Hall conductance is
$\sigma_{xy}^{\uparrow}=e^{2}/h$ ($\sigma_{xy}^{\downarrow}=-e^{2}/h$ ).
The whole system then has edge states with two spins 
propagating in the opposite direction. 
The whole system is T-symmetric, and the effective magnetic field
can be realized by the spin-orbit coupling. 

It is only the simplest special example. In general
the QSH allows a spin-orbit-coupling
term which mixes spins, without breaking T-symmetry.
The QSH phase is defined as a T-symmetric system 
which is gapful in the bulk and gapless in the edge.
Because of the T-symmetry, the edge states form Kramers pairs, consisting
of two states with opposite spins counter-propagating from each other.
The $Z_2$ topological number $\nu$ is used for distinguishing this
QSH phase from the usual insulator phase, namely the SH Iphase.
The $Z_2$ topological number can take only 
two different values for $\nu$: even or 
odd, and this number simply means whether the number of Kramers
pairs of edge states is even or odd.
If $\nu=$odd, the system is in the QSH phase, while
if $\nu=$even the system is in the insulating phase. This means that
the system  with an even number of Kramers pairs of edge states
is equivalent to a system with no gapless edge state. It follows because 
general perturbations preserving T-symmetry can open a gap 
in the edge states when $\nu=$even.
In contrast, for the QSH phase ($\nu=$odd), the edge
states remain gapless even in the presence T-symmetric 
perturbation, including nonmagnetic impurities and/or interaction,
as long as they are not too strong \cite{Wu05,Xu05}.

There are a number of equivalent expressions for the $Z_2$ topological number
\cite{Fu07a,Fu06a,Fu06c},
and we only explain the ones relevant for the subsequent discussions,
both in 2D and in 3D. We assume that the spectrum of the Hamiltonian has a gap,
within which the Fermi energy $E_F$ is located.

First we explain the 2D case.
For I-asymmetric systems,
the spectrum is doubly degenerate only at the four 
points $\bi{k}=\bi{k}_{i} \equiv \bi{G}/2$ ($i=1,2,3,4$), 
and non-degenerate at other 
points.
In such systems, the $Z_2$ topological number $\nu$ 
is determined as
\begin{equation}
(-1)^{\nu}=\prod_{i=1}^{4}\delta_{i},
\end{equation}
where \begin{equation}
\delta_{i}=\frac{\sqrt{{\rm det}[w(\bi{k}_{i})]}}{{\rm Pf}[w(\bi{k}_{i})]}
=\pm 1.
\label{eq:pfaffian}
\end{equation}
Here $w(\bi{k})$ is a unitary matrix with elements given by
$w_{mn}(\bi{k})=\langle u_{\bi{-k},m}|\Theta|u_{\bi{k},n}\rangle$, and 
$|u_{\bi{k},n}\rangle$ is the Bloch wavefunction of an 
$n$-th band whose eigenenergy lies below
$E_F$.
$\Theta$ is the time-reversal operator, represented as $\Theta=i\sigma_y K$ 
with $K$ being complex conjugation.
The branch of the square root of the determinant is so chosen 
that the wavefunctions (including their phases) 
are continuous in the whole Brillouin zone.

On the other hand, in I-symmetric systems, the formula simplifies 
drastically; there is no need to calculate
the phases of the wavefunctions for the whole Brillouin zone, 
which is advantageous for numerical 
calculation. It is given by
\begin{equation}
(-1)^{\nu}=\prod_{i=1}^{4}\delta_{i}, \ \ 
\delta_{i}=\prod_{m=1}^{N}\xi_{2m}(\bi{k}_{i}),
\label{eq:parity}
\end{equation}
where 
$\xi_{2m}(\bi{k}_i)$ ($=\pm 1$) is the parity eigenvalue of the 
Kramers pairs at each of these
points, and 
$N$ is the number of Kramers pairs below $E_F$.

In 3D, there are four 
$Z_2$ topological numbers written as 
$\nu_{0};(\nu_{1}\nu_{2}\nu_{3})$ \cite{Moore07,Fu07a}, given by
\begin{equation}
(-1)^{\nu_{0}}=\prod_{i=1}^{8}\delta_{i}, \ \
(-1)^{\nu_{k}}=\prod_{n_k=1; n_{j\neq k}=0,1}\delta_{i=(n_1n_2n_3)},
\label{eq:Z2-3D}\end{equation}
where $\delta_{i=(n_1n_2n_3)}$ ($=\pm 1$) is defined 
for the wavevector $\bi{k}_{i}=\frac{1}{2}(n_{1}\bi{b}_{1}+
n_{2}\bi{b}_{2}+n_{3}\bi{b}_{3})$ ($n_i=1,2,3$) and $b_{k}$ ($k=1,2,3$) are
the primitive vectors of the reciprocal lattice.  
These eight wavevectors satisfy $\bi{k}_{i}=-\bi{k}_{i}$
(mod $\bi{G}$).
These topological numbers in 3D determine the topology of the 
surface states for arbitrary crystal directions \cite{Fu07a}.
We note that among the four $Z_2$ topological numbers in 3D, 
only $\nu_{0}$ is robust against nonmagnetic impurities, 
while the others ($\nu_k$ ($k=1,2,3$)) are meaningful only 
for a relatively clean sample \cite{Fu07a}.

\subsection{Phase transitions between the quantum spin Hall and 
insulating phases}
The problem of our interest in this paper is how the $Z_2$ topological
number changes with a change of an external parameter.
For the I-symmetric systems it is easier to consider;
because it is the product of the parity, it can change when the valence
band and the conduction band with opposite parities 
touch and exchange their roles. 
On the other hand, in I-asymmetric systems
it is not obvious from (\ref{eq:pfaffian}) how
the $Z_2$ topological number changes.
For 2D, this was studied in the previous paper \cite{Murakami07a},
using the homotopy characterization of 
the $Z_2$ topological number in \cite{Moore07}. 
It is defined in the similar way as the Chern integer \cite{Moore07},
but with some modification.
From this definition, it follows that the $Z_2$ topological number in 2Dcan 
change when the valence and conduction bands touch 
each other at some $\bi{k}=\pm \bi{k}_{0}$.
(The band touching occur simultaneously at $\bi{k}=\pm \bi{k}_{0}$
because of the T-symmetry.) 
In 2D, the effective Hamiltonian 
in the vicinity of the phase transition reduces to 
\begin{equation}
{\cal H}=E_{0}(m,k_x,k_y)\pm (m-m_0)\sigma_z
+(k_x-k_{x0})\sigma_x+(k_y-k_{y0})\sigma_y
\label{eq:Weyl}
\end{equation}
after unitary and scale 
transformations, where $m$ is an external parameter which controls
the phase transition \cite{Murakami07a}. 
Equation (\ref{eq:Weyl}) describes the simplest and general 
case for the band crossing, which occurs at $(m,k_x,k_y)=(m_0,k_{x0},
k_{y0})$.

This story should be modified when the spatial dimension is three, 
especially for the I-asymmetric systems.
This is because (\ref{eq:Weyl}) cannot
accomodate
the four parameters $m$, $k_x$, $k_y$, and $k_z$, in contrast 
with the 2D counterpart (\ref{eq:Weyl}). In the following we
answer this question, by finding that there should lie a
gapless phase between the two gapped phases. This gapless phase is
a topological phase in the following sense. The phase transition 
is governed by monopoles, namely band-crossing 
between two non-degenerate
bands. Such monopoles are topological objects 
which can appear or dissappear not 
by itself,
but by creation/annihilation of a pair of 
a monopole and an antimonopole.

We first explain the framework for describing the phase transition in 3D.
As in the previous paper \cite{Murakami07a}, for the purpose of
describing generic phase transitions between the QSH and the insulating
phases, we consider a {\it single} parameter $m$ which controls the phase
transition. This parameter $m$ may be considered as externally 
controllable, and by changing this parameter the system undergoes 
the phase transition. 
At the phase transition the $Z_2$ topological number must change,
which necessitates the closing of the gap at some wavevector 
$\bi{k}$. 
There are various kinds of band crossing when we introduce
a number of parameters; nevertheless, to study the phase transition,
we restrict 
ourselves to ``generic'' band crossing, and exclude band crossing 
achieved only by tuning more than one parameters.

To be specific, we consider a Hamiltonian matrix
\begin{equation}
H(\bi{k})=\left(\begin{array}{cc}
h_{\uparrow\uparrow}(\bi{k})& h_{\uparrow\downarrow}(\bi{k})\\
h_{\downarrow\uparrow}(\bi{k})& h_{\downarrow\downarrow}(\bi{k})
\end{array}\right),\label{eq:Hamiltonian}
\end{equation}
where $\bi{k}=(k_x,k_y,k_z)$.
We assume that the spectrum of the Hamiltonian has no extra 
degeneracies other than those imposed by symmetry.
We also assume that the Fermi energy $E_F$ 
lies within a gap of the Hamiltonian.
The T-symmetry implies,
\begin{equation}
H(\bi{k})=\sigma_y  H^{T}(-\bi{k})\sigma_y,
\label{time-reversal}\end{equation}
i.e.\ $h_{\uparrow\uparrow}(\bi{k})
=h_{\downarrow\downarrow}^{T}(-\bi{k})$,
$h_{\uparrow\downarrow}(\bi{k})
=-h_{\uparrow\downarrow}^{T}(-\bi{k})$,
$h_{\downarrow\uparrow}(\bi{k})
=-h_{\downarrow\uparrow}^{T}(-\bi{k})$.
The Kramers theorem guarantees that 
the band structure of such T-symmetric spin-$1/2$ system is
symmetric with respect to $\bi{k}\leftrightarrow -\bi{k}$.
For the respective cases considered, it suffices to
choose the dimension of the Hamiltonian matrix
to be the number of states involved in band-crossing.

In 3D, as well as in 2D, there is another symmetry which is crucial 
for the nature of the phase transition: the inversion (I-)symmetry.
It is because it is the only symmetry beside the T-symmetry which 
transforms between $\bi{k}$ and $-\bi{k}$ for all $\bi{k}$.
We only consider these two (T- and I-)symmetries in this paper.
Various kinds of band crossings
found in higher point-group symmetries may be considered as 
degenerate cases of the generic cases considered here.
From the Kramers theorem for T-symmetric systems, 
the spectrum is doubly degenerate at the eight points 
$\bi{k}=\bi{k}_{i}=\bi{G}/2$ 
($i=1,\cdots,8$) in 
I-asymmetric systems, and 
is so for every $\bi{k}$ in I-symmetric systems.

\begin{figure}[h]
\includegraphics[scale=0.6]{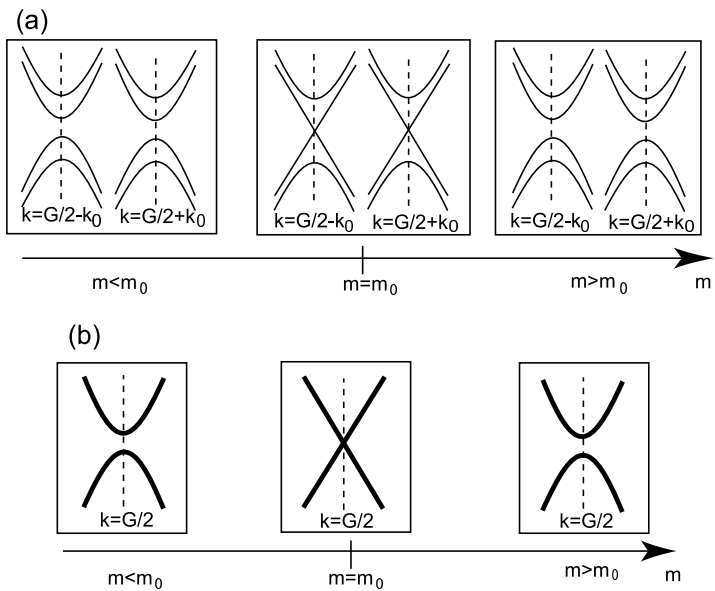}
\caption{Phase transition in 2D between 
the quantum spin Hall (QSH) and insulating phases
 for (a) inversion-asymmetric and 
(b) inversion-symmetric cases. In the case (b) all the states are
doubly degenerate.}
\label{fig:degeneracy2D}\end{figure}

Let us explain the phase transitions in 2D as obtained in the previous paper
\cite{Murakami07a}.
In I-asymmetric systems 
the band crossings occur at $\bi{k}=\pm \bi{k}_{0}\neq \bi{G}/2$, 
between non-degenerate bands (\fref{fig:degeneracy2D}).
Because of the T-symmetry, band crossing occurs simultaneously at $\bi{k}=
\pm \bi{k}_{0}$, at a single point $m=m_{0}$.
On the other hand, in I-symmmetric systems, the band crossing
occur at $\bi{k}=\bi{k}_{i}=\bi{G}/2$ between two doubly-degenerate bands.
The two bands should 
have an opposite parity, and their parities are exchanged
at the band crossing \cite{Murakami07a}.
In the following we show that the phase transition in the I-symmetric systems
are similar between 2D and 3D, whereas in the I-asymmetric systems
they are quite different.

\subsubsection{I-asymmetric systems}
In 3D, in contrast with the 2D case, band crossing at 
$\bi{k}=\pm \bi{k}_{0}\neq \bi{G}/2$ 
cannot lead to phase transion.
The reason is the following. 
The energy bands for the I-asymmetric  systems are nondegenerate 
for $\bi{k}\neq\bi{G}/2$. A crossing 
of two such energy bands has codimension three; namely, by tuning 
three parameters one can make two bands 
degenerate \cite{vonNeumann29,Herring37b}. 
To see this, let us consider $2\times 2$ Hamiltonian matrix
\begin{equation}
H=\left(
\begin{array}{cc}
a & c\\
c^{*} & b
\end{array}
\right),
\end{equation}
where
$a$, $b$ are real functions of $\bi{k}$ and $m$, and $c$ is 
a complex function of $\bi{k}$ and $m$. 
A necessary condition for the two eigenvalues to be identical 
consists of three conditions $a=b$, ${\rm Re}c=0$ and ${\rm Im}c=0$,
i.e. the codimension is 3 \cite{vonNeumann29}.
These three conditions determine a curve 
in the four-dimensional space $(m,k_x,k_y,k_z)$.
Thus for generic $m$ there
will be in general a point (or points) $\bi{k}$ where the eigenvalues 
are degenerate. When $m$ is changed continuously the $\bi{k}$
point moves in the $\bi{k}$ space, and the system 
remains gapless.
It will be revealed later how the system can open a gap and run into 
either the QSH or the insulating phases.

On the other hand, for $\bi{k} = \bi{G}/2$, band crossing cannot occur
in general. 
At the points $\bi{k} = \bi{G}/2$,
the spectrum 
is doubly degenerate, and the codimension 
is five \cite{Avron88,Avron89}. It can be explicitly seen as follows.
As the number of states involved is four, we consider $4\times 4$ 
Hamiltonian matrix with the constraint (\ref{time-reversal}).
It leads to a result
\begin{equation}
{H}(\bi{k}=\bi{k}_{i})=E_{0}+\sum_{i=1}^{5}a_{i}
\Gamma_{i}
\label{eq:asym-Gamma-i}
\end{equation}
where $a_{i}$'s and $E_{0}$ are real, and
$\Gamma_{1}=1\otimes\tau_{x}$, $\Gamma_{2}=\sigma_{z}\otimes\tau_{y}$, 
$\Gamma_{3}=1\otimes\tau_{z}$, 
$\Gamma_{4}=\sigma_{y}\otimes\tau_{y}$, and  
$\Gamma_{5}=\sigma_{x}\otimes\tau_{y}$.
Its eigenenergies are given by 
$E_{0}\pm\sqrt{\sum_{i=1}^{5}a_{i}^{2}}$. 
The two (doubly-degenerate) bands will touch when 
$a_{i}=0$ for $i=1,\cdots,5$, 
which are not satisfied by tuning only one parameter $m$. (Note that
the wavenumber $\bi{k}$ is fixed here and cannot be changed.)
Thus it is impossible 
to control the bands to touch at $\bi{k}=\bi{G}/2$ 
by tuning a single parameter $m$.

\subsubsection{I-symmetric systems}
In I-symmetric systems, the energies 
are doubly degenerate for every $\bi{k}$ by the Kramers theorem. 
The phase transition occurs when the gap 
between the two doubly-degenerate bands closes at some $\bi{k}$.
Because there are four states involved, 
we consider the 4$\times$4 Hamiltonian 
matrix $H(\bi{k})$.
We impose the I-symmetry as
\begin{equation}
H(-\bi{k})=PH(\bi{k})P^{-1}, \ u(-\bi{k})=Pu(\bi{k}),
\label{eq:H-inversion}
\end{equation}
where $P$ is a unitary matrix independent of $\bi{k}$, 
and $u(\bi{k})$ is the periodic part 
of the Bloch wavefunction: $\varphi_{\bi{k}}(\bi{r})
=u(\bi{k})e^{i\bi{k}\cdot\bi{r}}$.
After a judicious unitary transformation, all cases reduce to 
\begin{equation}
P=\left(
\begin{array}{cc}
P_{\uparrow}&\\
&P_{\downarrow}\end{array}
\right),\ \ 
P_{\uparrow}=P_{\downarrow}={\rm diag}(\eta_{a},\ \eta_{b})
\end{equation}
 without losing generality. $\eta_a$ and $\eta_b$ represent the
parity eigenvalues of the atomic orbitals involved.

The band crossings are
different 
for
$\eta_a=\eta_b$ and 
$\eta_a=-\eta_b$ in 3D, as is similar to 2D \cite{Murakami07a}.
When $\eta_{a}=\eta_{b}=\pm 1$, the generic Hamiltonian 
becomes
\begin{equation}
{H}(\bi{k})=E_{0}(\bi{k})+\sum_{i=1}^{5}a_{i}(\bi{k})
\Gamma_{i}
\label{eq:sym-same}
\end{equation}
where $a_{i}$'s and $E_{0}$ are real even functions of $\bi{k}$.
On the other hand, when $\eta_{a}=-\eta_{b}=\pm 1$, 
the Hamiltonian reads,
\begin{equation}
{H}(\bi{k})=E_{0}(\bi{k})+a_{5}(\bi{k})\Gamma'_{5}+\sum_{i=1}^{4}
b^{(i)}(\bi{k})
\Gamma'_{i}
\label{eq:sym-different}
\end{equation}
where $E_0(\bi{k})$ and $a_{5}(\bi{k})$ are even functions of $\bi{k}$,
$b^{(i)}(\bi{k})$ are odd functions of $\bi{k}$.
The matrices $\Gamma'_{1}=\sigma_{z}\otimes
\tau_{x}$,
$\Gamma'_{2}=1\otimes\tau_{y}$,
$\Gamma'_{3}=\sigma_{x}\otimes\tau_{x}$,
$\Gamma'_{4}=\sigma_{y}\otimes\tau_{x}$,
and $\Gamma'_{5}=1\otimes\tau_{z}$ form the Clifford algebra.
Therefore, for generic point $\bi{k}=\bi{k}_{i}\neq\bi{G}/2$, 
In both cases, $\eta_a=\eta_b$ and $\eta_a=-\eta_b$, 
the codimension is five, which exceeds the number of tunable 
parameters ($m$, $k_x$, $k_y$, $k_z$).
Therefore, band crossing does not occur at a generic point
$\bi{k}$ with $\bi{k}\neq \bi{k}_i \equiv \bi{G}/2$,

On the other hand, at the high-symmetry points  $\bi{k}=\bi{k}_{i}
=\bi{G}/2$. The number of parameters to
achieve degeneracy is five for $\eta_a=\eta_b$ (in (\ref{eq:sym-same}))
while it is one for $\eta_a=-\eta_b$ (in (\ref{eq:sym-different})).
Because the wavenumber is fixed
 $\bi{k}=\bi{k}_{i}
=\bi{G}/2$, there is only one changeable parameter $m$. Thus only
when $\eta_a=-\eta_b$, the two-doubly degenerate bands touch at 
$\bi{k}=\bi{k}_{i}=\bi{G}/2$.
This situation is similar to the case in 2D \cite{Murakami07a}.

Thus we have seen that in I-symmetric 3D systems, 
the phase transition can occur at a single 
value of the parameter $m=m_0$. At one side (e.g. $m<m_0$), the system
is in the QSH while the other side (e.g. $m>m_0$) is in the insulating phase.
Meanwhile, the phase transition 
at a single value of $m$ is in general impossible in I-asymmetric cases.
Here we encounter a question. How is the phase transition in the 
I-symmetric 3D system modified when some perturbation breaks
I-symmetry.
Because the two sides ($m\ll m_0$ and $m\gg m_0$) belong to the 
different phases,
there should be a phase transition in between.

\begin{figure}[h]
\includegraphics[scale=0.6]{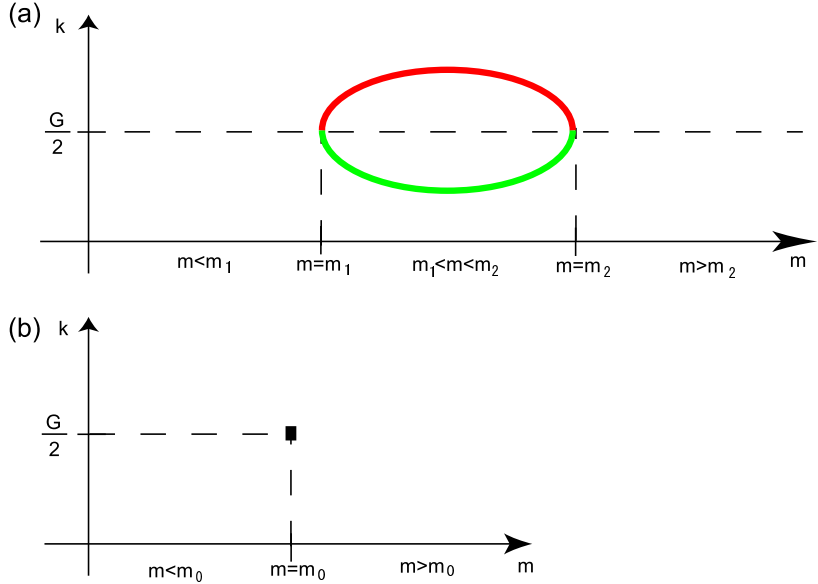}
\caption{Location of the the gapless points by changing the 
external parameter $m$ in (a) I-asymmetric systems and (b) I-symmetric
systems. In (a) the green and the red denotes trajectories
of the monopole and antimonopole, respectively.}
\label{fig:monopole}\end{figure}

The answer is the following. Instead of a phase transition 
occuring at the single value of the parameter $m$, there appears 
a finite region of $m$ where the system remains gapless. 
As was discussed previously, the band crossing in the I-asymmetric
system cannot occur at a single value of $m$. 
The gapless points in the four-dimensional $m$-$\bi{k}$ space
forms a one-dimensional manifold (i.e. a curve). Thus, the only 
way to close the 
gap by changing $m$ is to make a pair of 
gapless points, as shown in \fref{fig:monopole}. 

To see the behavior of such gapless points, 
we note that each gapless point carries a topological number. 
Such gapless point, sometimes called a diabolical point, is regarded as 
a monopole in the $\bi{k}$ space \cite{Volovik87,Volovik01,Murakami03}. 
Indeed, the Bloch wavefunctions with nondegenerate spectrum 
can be associated with a U(1) gauge structure in the $\bi{k}$ space;
\begin{eqnarray}
\bi{A}_{n}(\bi{k})&=&-i\langle \bi{k}n|\nabla_{\bi{k}}|\bi{k}n\rangle,\\
\bi{B}_{n}(\bi{k})&=&\nabla_{\bi{k}}\times \bi{A}_{n}(\bi{k}),\\
\rho_{n}(\bi{k})&=&\frac{1}{2\pi}\nabla_{\bi{k}}\cdot \bi{B}_{n}(\bi{k})
\end{eqnarray}
The quantity $\rho(\bi{k})$ is called a monopole density. 
It vanishes when the $n$-th band is not degenerate with 
other bands. At the band crossings between the $n$-th band
and other bands, $\rho(\bi{k})$ has a $\delta$-function singularity;
$\rho_{n}(\bi{k})=\sum _{l}q_{ln}\delta(
\bi{k}-\bi{k}_{ln})$ where $q_{ln}$ is an integer called
a monopole charge. 
For example, 
for the band crossing at $\bi{k}=\bi{k}_{0}$ with 
linear dispersion (Weyl fermion)
\begin{equation}
{\cal H}=E_0(\bi{k})+\sum_{i=1}^{3}f_i(\bi{k})\sigma_i
\label{eq:Weyl2}
\end{equation}
where $f_i(\bi{k}=\bi{k}_{0})=0$, the monopole charge 
for the lower band
is ${\rm sgn}({\rm det}(\frac{\partial f_i}{\partial k_j})_{ij})|_{\bi{k}=\bi{k}_{0}}$
($=\pm 1$). 

In the present system, a pair of a monopole (charge $q=1$) and an antimonopole 
(charge $q=-1$) is
created at $m=m_1$, $\bi{k}=\bi{k}_{i}=\bi{G}/2$ 
when $m$ is increased, and the system becomes gapless.
When $m$ is increased further, the monopole ($\bi{k}=\bi{k}_{+}$) 
and the antimonopole $(\bi{k}=\bi{k}_{-}$)
moves in the $\bi{k}$ space, while the T-symmetry imposes that 
$\bi{k}_{-}=-\bi{k}_{+}$.
This system can open a gap again only when the monopole and antimonopole
annihilate together. The annihilation can occur only at the points
with $\bi{k}=\bi{k}_{i}=\bi{G}/2$, again by the T-symmetry
(\fref{fig:monopole}). 
Thus the overall feature of the phase transition is
schematically expressed as in \fref{fig:degeneracy3D}.

\subsubsection{Summary for the phase transition in 3D}
\begin{figure}[h]
\includegraphics[scale=0.6]{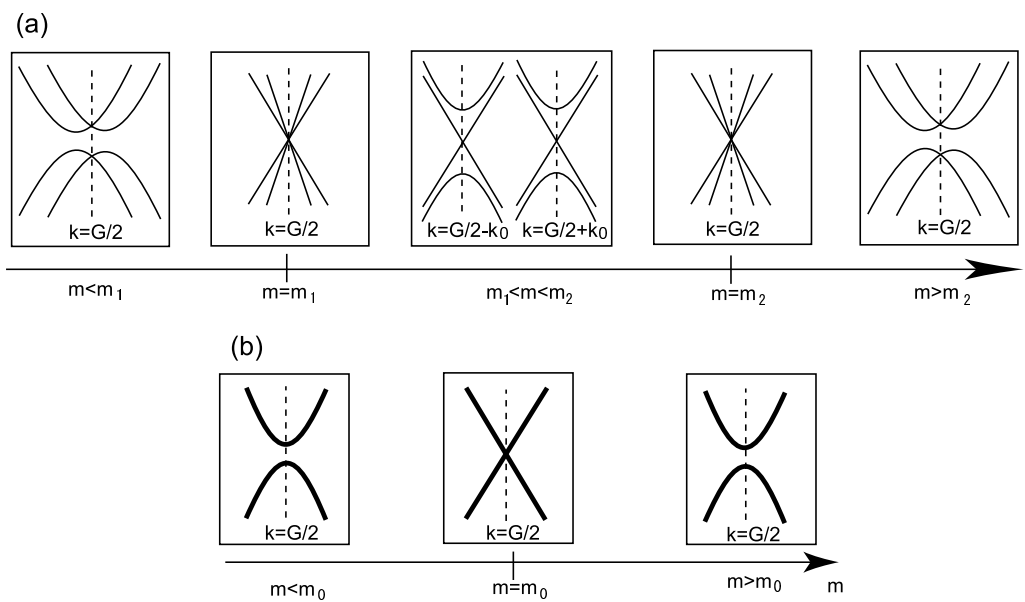}
\caption{Phase transition in 3D between 
the quantum spin Hall (QSH) and insulating phases
 for (a) inversion-asymmetric and 
(b) inversion-symmetric cases. In the case (b) all the states are
doubly degenerate.}
\label{fig:degeneracy3D}\end{figure}

\begin{figure}[h]
\includegraphics[scale=0.6]{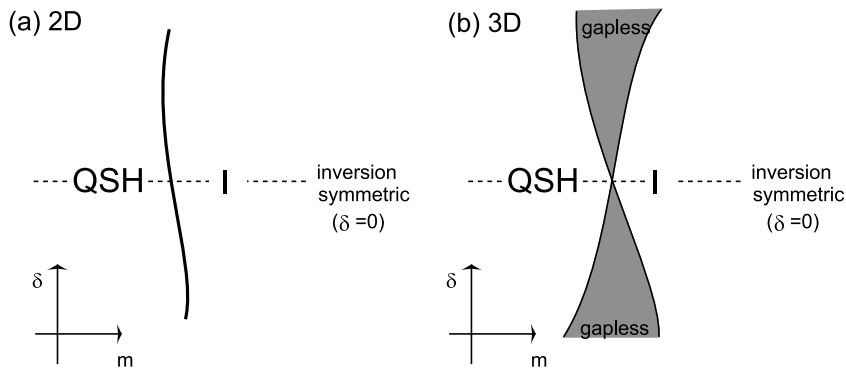}
\caption{Phase diagram for 
the quantum spin Hall (QSH) and ordinary insulating (I)
phases for (a) 2D and (b) in 3D.
$m$ is a control parameter which drives the phase transition, and
$\delta$ represents a parameter which describes the breaking of 
inversion symmetry. $\delta=0$ is the case with inversion symmetry.}
\label{fig:phase-diagram}\end{figure}

To summarize, 
the overall feature of the band crossing is schematically shown in 
\fref{fig:degeneracy3D}. The overall phase diagram is schematically 
shown in 
\fref{fig:phase-diagram}, in a plane of the control parameter
$m$ and another parameter $\delta$ 
representing an inversion-symmetry breaking. As we have seen,
when the I-symmetry is broken, the topological gapless phase
appears between the two phases which are gapful in the bulk.
The difference of the $Z_2$ topological
numbers $\nu$ between the two sides of the phase transition can also
be calculated as in the 2D case \cite{Murakami07a}. 
When the band crossing occurs at $\bi{k}_{i}=\frac{1}{2}(n_{1}\bi{b}_{1}+
n_{2}\bi{b}_{2}+n_{3}\bi{b}_{3})$,
the factor $\delta_{i=(n_1n_2n_3)}$ in (\ref{eq:Z2-3D}) changes sign, and 
some of the four $Z_2$ topological numbers, $\nu_{0};(
\nu_1\nu_2\nu_3)$ in (\ref{eq:Z2-3D}), change accordingly.
This applies to the I-symmetric systems.
The I-asymmetric cases are similarly treated because it can be
associated with the I-symmetric cases with perturbation.

\section{Towards materials search for quantum spin Hall systems}
\subsection{Bismuth thin film and 2D quantum spin Hall system}
In \cite{Murakami06b}, the bilayer bismuth is studied as 
a candidate for the 2D QSH phase.
From the 2D bilayer tight-binding model, truncated from the 3D tight-binding 
model \cite{Liu95}, the bilayer bismuth is proposed to be the 2D QSH
system
\cite{Murakami06b}.
Bilayer antimony is studied in a similar way, 
and it is the ordinary insulator.
It is predicted from the calculation of $Z_2$ topological number and 
from a band structure calculation for the geometry with edges (i.e. 
the strip geometry). The $Z_2$ topological number is calculated in 
\cite{Murakami06b} by the Pfaffian of the matrix for the
time-reversal operator proposed in \cite{Kane05a}. 

The calculation of the Pfaffian involves fixing of phases of the 
wavefunction as an analytic function of the wavenumber ${\bf k}$, 
which is numerically a challenging problem, even for a simplified 
model presented in \cite{Murakami06b}. 
It can be tackled by discretizing the ${\bf k}$ space and counting 
the vortex of the Pfaffian matrix \cite{Fukui07a,Fukui06b}. 
Instead, for I-symmetric systems, 
the method of calculating parity eigenvalues 
(\ref{eq:parity}) proposed in \cite{Fu06c} is much easier. 
For the bilayer bismuth, we checked that this method
leads us also to the same conclusion that the $Z_2$ topological number
is odd and nontrivial, and it is in the QSH phase.

For the system to be the QSH phase, it should have a gap in the bulk.
The 3D bulk bismuth is semimetallic, and have a small band overlap 
between the conduction and the valence band, while the direct 
gap is finite for all wavenumbers.
By making it into thin film, the perpendicular motion is quantized 
and tends to open a band gap. 
In earlier theoretical estimate \cite{Lutskii65,Sandomirskii67}
and experiments \cite{Hoffman93,Hoffman95} (see also \cite{Chu95})
shows that in thin-film bismuth thinner than $\sim 30{\rm nm}$
becomes an insulator.
However,
in recent experiments by Hirahara {\it et al.} \cite{Hirahara06} 
the angle-resolved photoemission spectra (ARPES) 
the ultrathin films of 10 bilayers
are measured. In their experimental data, as opposed to the 
earlier theories and experiments, the system remains semimetallic.
It casts us a question that the critical thickness might be 
much smaller than the earlier estimate
\cite{Lutskii65,Sandomirskii67,Hoffman93,Hoffman95}.

\subsection{Bismuth and 3D quantum spin Hall system}
In \cite{Fu06c} it was suggested that the 3D Bi has $\nu=$even, 
while the 3D Sb has $\nu=$odd.
This looks opposite to the case of the 2D. 
To resolve this,  
numerical analysis was performed in \cite{Fukui06b}
by artificially changing the interlayer hopping
multiplied by a factor $f$ ($0<f<1$).
It was found that for Bi, the regions
$0\leq f< 0.223$, $0.223<f<0.993$, and $0.993<f\leq 1$
are the phases 0;(111), 1;(111), and 0;(000), respectively, 
whereas for Sb the regions
$0\leq f< 0.54$ and $0.54<f\leq 1$ are the 
phases 0;(000) and  1;(111).
This calculation was done by the calculation of the Pfaffian matrix,
while this result was checked by the parity analysis proposed in \cite{Fu06c}.
The phase transitions by changing $f$ 
are described by our theory developed in the previous section.
The band crossing which accompanies the phase transition occurs
for Bi at $(0,0,0)$ for $f=0.223$, and at $(\pi,0,0)$, $(0,\pi,0)$,
and $(0,0,\pi)$ for $f=0.993$. 
For Sb, the band crossing occurs at $(\pi,\pi,\pi)$ for $f=0.54$.
These systems are I-symmetric. According to our theory, 
the band crossing occurs only at 
the single value of the control parameter $f$, rather than having a 
gapless phase in between. 

In reality, both Bi and Sb are semimetals, not insulators. 
The $Z_2$ topological numbers are defined in the 3D, by assuming
that the band overlap is lifted by some perturbation, thereby 
the bands below the gap are regarded as the ``valence band'' which 
enters in the definition of the $Z_2$ topological numbers.
This is possible because there is a direct gap in every $\bi{k}$.
In this sense, although the 3D $Z_2$ topological number $\nu_0$ 
defined as such
is odd (nontrivial) in Sb, it is not the QSH phase, because there is
no gap. If one can open a gap in the 3D Sb by some external perturbation,
it becomes the QSH phase.

Nevertheless, because the $Z_2$ topological numbers can be defined in Bi and Sb
in the above sense, they manifest themselves in the spectrum of surface states.
The 3D bulk Bi and Sb has the topological numbers 0;(000) and 1;(111),
from which expected 
topology of the Fermi surface of the surface states 
can be
easily sorted out for various directions of crystal surface
\cite{Fu06c}. 
The results can be compared with the experiments on 
the angle-resolved photoemission spectroscopy (ARPES) 
for Bi (for example, \cite{Koroteev04}) and for Sb \cite{Sugawara06}.
To interpret these experiments to see whether it matches the
prediction from the $Z_2$ topological number, we need to
separate the Fermi surface of bulk states and that of surface
states, which is not trivial experimentally.

To find a clear experimental manifestation
for the nontrivial $Z_2$ topological number, we need anyway open 
a gap in the bulk. One example is Bi$_{1-x}$Sb$_{x}$ ($0.07<x<0.22$),
and 
ARPES experiments on this doping region is called for.
To our knowledge, \cite{Hoechst05} is the only experimental report
in this doping region. Nonetheless, this experiment is limited only in the
vicinity of the $\bi{k}=0$, and not sufficient to examine its
topological phase from the ARPES data. A data on ARPES in Bi$_{1-x}$Sb$_{x}$ 
($0.07<x<0.22$) covering the whole surface Brillouin zone would be
a clear evidence for the QSH phase.
Thus, there is no clear experimental evidence 
for the QSH phase as yet, whereas
it looks well within our reach, and a good candidate is strongly called for.

\subsection{Criterion for searching quantum spin Hall systems}
Apart from the time-reversal symmetry, there are only two conditions
for the QSH phases.
\begin{enumerate}
\item the bulk is a band insulator
\item the $Z_2$ topological number is odd
\end{enumerate}
The first condition is clear, while the second requires calculation. 
At this stage we need some strategy to search 
among the vast number of nonmagnetic insulators.

To find out the strategy, let us begin 
with a system without spin-orbit coupling, 
and switch on the spin-orbit 
coupling gradually.
Insulators without the spin-orbit coupling have
a trivial (i.e. even) $Z_2$ topological number.
To reach the QSH phase, the gap should once closes
in switching on the spin-orbit coupling, thereby
the system should undergo a phase
transition.
This phase transition 
should be described within our theory 
developed in the previous section.

Therefore, the QSH phase near the phase transition may have the direct gap
at $\bi{k}=\bi{k}_{i}=\bi{G}/2$, as a trace of the phase transition.
When this phase transition occurs similtaneously at some points in 
the Brillouin zone, the changes of the $Z_2$ topological number add
together. In this sense, when the material has an even number of equivalent
points
for the direct gaps which close at the phase transition, 
the $Z_2$ topological number $\nu_0$ 
does not change at the transition, and the system remains 
an ordinary insulator.
The example is the 4 equivalent $L$ points in 
PbTe \cite{Fu06c}, and PbTe is indeed in the ordinary insulator 
 phase.

Therefore it is desirable to have direct gaps at an odd number of 
points in the Brillouin zone, such as the $L$
points in Bi$_{1-x}$Sb$_{x}$ \cite{Fu06c}, or  the $\Gamma$ point
in generic crystals.
We note that this is not a necessary condition, but can be a reasonable
guideline for searching candidate materials among nonmagnetic insulators.

It is also necessary to see whether the gap in the material considered
is ``after'' or ``before'' the phase transition, when one
turns on the spin-orbit coupling gradually.
In other words, for the QSH phase, the 
gap should be originated from the spin-orbit coupling.
This statement is somewhat vague; to make this more transparent,
the author suggested that the 
susceptibility can be a measure to see whether the 
gap is of the spin-orbit nature \cite{Murakami06b}. 
For example, 
bismuth are strongly diamagnetic, because of
the inter-band matrix elements between the conduction and the
valence bands due to the spin-orbit couping.
This class of materials, when gapped, are good candidates for
the QSH phases. 
The criteria discussed so far will be useful for finding good
candidates for the QSH phase.

\section{Conclusions and Discussions}
In the present paper we studied the phase transition between the QSH and 
the insulating phases in 3D. In contrast to the 2D systems, 
in the 3D inversion-asymmetric cases there is a gapless phase between
the two phases. 
This gapless phase originates from the topological nature of the monopoles
(band-crossing points) in 3D, which are responsible for
the phase transition.
Furthermore, the gap closing occurs only at the high-symmetry points
$\bi{k}=\bi{G}/2$, which is also in contrast with 2D cases.

So far we have given topological argument in terms of the Bloch wavefunctions.
One may suspect that the scenario may become invalid in the presence 
of interaction or disorder, thereby the topological order may be obscured.
We give here qualitative argument 
that our scenario is robust against small perturbations.
Our argument is based on the fact that the $Z_2$ topological number 
(for gapped systems) 
can be defined even in the presence of interaction and disorder \cite{Fu06c},
by use of an analogue of 
the Laughlin's gedanken experiment \cite{Laughlin81}.
In disordered systems, the Bloch wavenumber $k_x$ becomes
ill-defined. 
We then think of folding the system to 
a ``ring'', periodic in the $x$ direction, and threading a flux $\Phi_x$ into
its hole. The flux $\Phi_x$ plays the role of $k_x$.
A similar procedure is taken for $k_y$ and $k_z$. The $Z_2$ topological number 
is defined by using $\Phi_i$ ($i=x,y,z$) instead of $k_i$.
Therefore, as long as interaction and disorder is relatively weak,
the bulk remains gapped and the $Z_2$ topological number remains
well-defined and idential with that in the clean, noninteracting systems.
As is related with this stability of the topological order, the 
gapless surface state in the 3D QSH phase 
remain gapless and show antilocalization behaviour 
in the presence of impurities.
It is in the symplectic universality class, 
and is similar to the honeycomb lattice without 
intervalley scattering \cite{Suzuura02}. 
For this reason, the 3D QSH phase is robust against impurities.

Based on these observations, we can now see that the gapless topological phase
($m_1<m<m_2$ in Figure \ref{fig:degeneracy3D}(a)) will survive in 
the presence of disorder and interaction.
Since the two sides of the topological gapless phase are 
the gapped phases ($m<m_1$ and $m>m_2$ in Figure \ref{fig:degeneracy3D}(a))
with different $Z_2$ topological numbers, there should 
lie a gapless phase in between, even in the presence of interaction and disorder. (If the topological gapless phase becomes 
gapped, it can no longer accompany a phase transition.)
Quantitative argument to see how this gapless phase is robust 
is involved and
is beyond the scope of the present paper,

A following remark is in order. A calculation for the
$Z_2$ topological number based on a simplified model  
requires some care, because the bands well below the Fermi energy 
might contribute to the $Z_2$ topological nubmer.
We present one example. In \cite{Murakami04c} we constructed
a 4-band tight-binding model for HgTe.
It is a zero-gap system, but by opening a gap with uniaxial pressure, 
it becomes a band insulator. The question here is whether it is the
simple insulator or the topological insulator (i.e. QSH).
In fact, from the parities of the bands below the Fermi energy $E_F$
from the four-band model in \cite{Murakami04c}, 
the $Z_2$ topological number $\nu$ turns out to be even. 
It disagrees with the result in \cite{Fu06c}.
The reason for the disagreement is the following.
This model only has four bands near the Fermi energy $E_F$,
 and the other bands 
well below $E_F$ are 
discarded. In HgTe, one doubly-degenerate band just below the Fermi energy
has a trivial $Z_2$ topological number, while the other discarded 
bands deep below the 
Fermi energy has a nontrivial $Z_2$ topological number; this leads to the
seemingly contradicting results. Thus in the 
calculation of the $Z_2$ topological number one should be careful
because it involves
all the bands below the Fermi energy $E_F$.

The lesson we can learn from 
the examples of bismuth \cite{Murakami06b} and other materials \cite{Fu06c}
is that the materials with odd $Z_2$ topological numbers are not rare.
In nature there may well exist systems with odd $Z_2$ topological number. 
The crucial difference between the two kinds of topological insulators, 
namely the quantum Hall (QH) 
and the QSH system, is that 
the QH systems require magnetic field as strong as several teslas, 
whereas the QSH systems does not require external fields to achieve
topological phases. This is a crucial difference.
To realize the QSH phase, the effective magnetic field produced from 
the spin-orbit coupling should be of some teslas, as expected from the 
analogy with the QH system. We see from the examples of bismuth 
\cite{Murakami06b} and other materials \cite{Fu06c} that the spin-orbit 
couping is strong enough in some materials. 
Thus it is natural 
to expect some materials in nature 
to be in the QSH phase, which would be of
great interest both theoretically and experimentally.

\ack
The author would like to thank 
Y.~Avishai, T. Hirahara, 
S.~Iso, N.~Nagaosa, M. Onoda, and S.-C. Zhang for fruitful 
discussions and helpful comments.
This research is supported in part 
by Grant-in-Aid and NAREGI Nanoscience Project 
from the Ministry of Education,
Culture, Sports, Science and Technology of Japan.  

\end{document}